# Active Vibration Control of Launch Vehicle on Satellite Using Piezoelectric Stack Actuator

Mehran Makhtoumi

*Abstract*— **Satellites are subject to various severe vibration during different phases of flight. The concept of satellite smart adapter is proposed in this study to achieve active vibration control of launch vehicle on satellite. The satellite smart adapter has 18 active struts in which the middle section of each strut is made of piezoelectric stack actuator. Comprehensive conceptual design of the satellite smart adapter is presented to indicate the design parameters, requirements and philosophy applied which are based on the reliability and durability criterions to ensure successful functionality of the proposed system, also fabrication process of the proposed piezoelectric stack actuator is discussed in detail. The coupled electromechanical virtual work equation for the piezoelectric stack actuator in each active strut is derived by applying D'Alembert's principle to calculate the consistent mass matrix, the stiffness matrix and the load vector using finite element approximation. Modal analysis is performed to characterize the inherent properties of the smart adapter and extraction of a mathematical model of the system. Active vibration control analysis was conducted using fuzzy logic control with triangular membership functions and acceleration feedback. The control results conclude that the proposed satellite smart adapter configuration which benefits from piezoelectric stack actuator as elements of its 18 active struts has high strength and shows excellent robustness and effectiveness in vibration suppression of launch vehicle on satellite.**

*Index Terms*— **Vibration Control, Piezoelectric, Fuzzy Logic Control, Launch Vehicle, Satellite, D'Alembert's Principle, Finite Element Method, Acceleration Feedback**

## Nomenclature

$\sigma_{ij}$ = stress vector

$C_{ij}^{E}$ = elasticity stiffness constant

$e_{ij}$ = piezoelectric stress coefficient

$E_m$ = vector of applied electric field

$D_m$ = vector of electric displacement

$\varepsilon_{ik}^{\sigma}$ = permittivity

$E_k$ = vector of applied electric field

$\varepsilon_i$ = strain vector

*Subscripts*

*PSA* = piezoelectric stack actuator

## I. INTRODUCTION

Launch vehicles are the source of various severe vibration during different phases of flight. Factors ranging from stage separation to propulsion ignition and shut down along with acoustic and aerodynamic forces all would cause dynamic excitation of launch vehicle structures. Propagation of these disturbances to the satellites could cause precision loss, damage or complete mission failure [1-3]. Since the spread of these dynamic loads occur via a structural path through the adapter to the satellite, thus developing a smart adapter which could actively control dynamic response of the structure is of the utmost importance.

Conventional payload adapters require developing a circular ring or conical shell structural configuration [4-7] which have gained the least vibration control efficiency. From among many advanced techniques to control vibration in space structures, piezoceramic actuated systems have resulted in the most reliable outcomes [8-12]. As for instance, active vibration control of acoustic and dynamic excitations by utilizing piezoceramic actuators on the cylindrical shell structure has been investigated by teams of scientists [13-15]. Previous researches concern modal methods application for active vibration control of shell structures along with piezoelectric actuator application for internal cavity noise control which mainly perform modal spectra for coupling between the cylindrical and internal acoustic cavity modes [16-19]. Numerical analysis of a simple cylindrical shell model with piezoelectric actuators shows that the large piezoelectric actuators would be more effective than small ones [20].

In order to achieve a successful approach that could fulfill the requirements of the active vibration control of launch vehicle on satellite, it is essential to propose a new device and systems which would benefit from an innovative structural configuration. Although the piezoelectric patched systems reported for active vibration control of shell structures in the literature are considered as an engineering marvel, in this study the aim is to stretch the piezoelectric vibration control nature to its maximum capacity by merging stack forms of piezoelectric material into a new satellite adapter configuration. In this study, the concept of satellite smart adapter is proposed to satisfy active vibration control of launch vehicle on satellite. The proposed smart adapter has 18 active struts in which the middle section of each strut is made of piezoelectric stack actuator. The main advantage of utilizing piezoelectric stack actuators is that they are characterized by their capacity to operate in high stress, voltage, and temperature environments. Because of these properties, piezoelectric stacks are the most suitable and

Mehran Makhtoumi is an Aerospace Structures Design Engineer and currently is with Universidad Politecnica de Cataluña, Barcelona, Spain
E-mail: mehran.makhtoumi@upc.edu

effective electromechanical actuators which could be used as elements of the proposed satellite smart adapter.

The contributions of this study include:

- Satellite smart adapter concept; a comprehensive conceptual design to indicate the design parameters, requirements and philosophy applied which are based on the reliability and durability criterions to ensure successful functionality of the proposed system.
- Dynamic model formulation; to derive the coupled electromechanical virtual work equation for the piezoelectric stack actuator in each active strut.
- Modal analysis; to characterize the inherent properties of the satellite smart adapter.
- Control design; to outline the fuzzy logic control algorithm for active vibration control of the system.
- Results and discussion; to illustrate and analyze the controlled responses of the smart adapter system.

Also, it is worth to note that the future investigation of this work is dedicated for comprehensive finite element analysis of each active strut with solving correspondent partial differential equations presented in this study by utilizing efficient numerical techniques which are well addressed in the references [21-24].

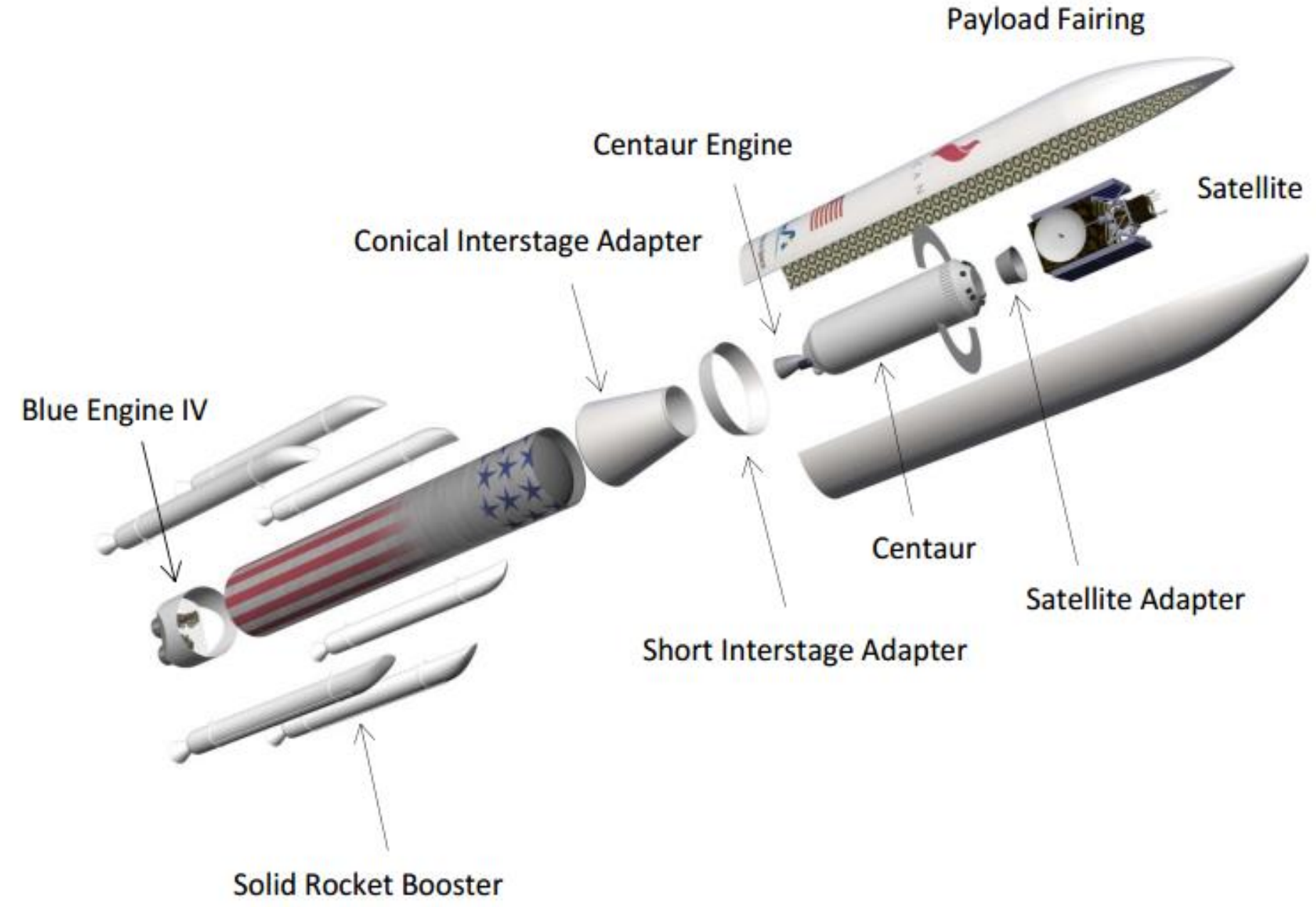

Fig. 1. Vulcan launch vehicle with conical satellite adapter

## II. SATELLITE SMART ADAPTER CONCEPT

The main aim is to propose a satellite smart adapter concept which could replace conventional conical shell adapter structure on the Vulcan launch vehicle shown in Fig. 1.

The smart adapter concept consists of two interfaces; one at the bottom to make joint with Centaur and the other at the top for satellite settlement connection. These interfaces are connected by 18 active struts together which benefit from fork head type connections, the overall model is illustrated in Figure 2 and active strut in Figure 3. From trial and errors, it is concluded that a large number of the active struts will only increase the system complexity and it is not cost-effective thus the best model will utilize 18 active struts. Also, a fail-safe design criterion has been applied to the system which in case of any mechanical or electrical failure would enable the smart adapter to act as a passive adapter. In the proposed smart adapter, each strut is divided into three sections and actuator part is located at the middle.

Piezoelectric actuators are produced in two forms of patch and stack, since piezoelectric stack actuators have the ability to withstand high pressure and show the highest stiffness of all piezo-actuator designs thus piezoelectric stacks are placed on the active part of each strut in the middle sections which the PIC-151 piezo-material is proposed to be used. The reason for selecting PIC-151 piezo-material is that they are characterized by very high operating voltage (up to 1000 volt), extreme reliability (> $10^9$cycles) and high force generation (up to 80 kN) which makes them extremely durable.

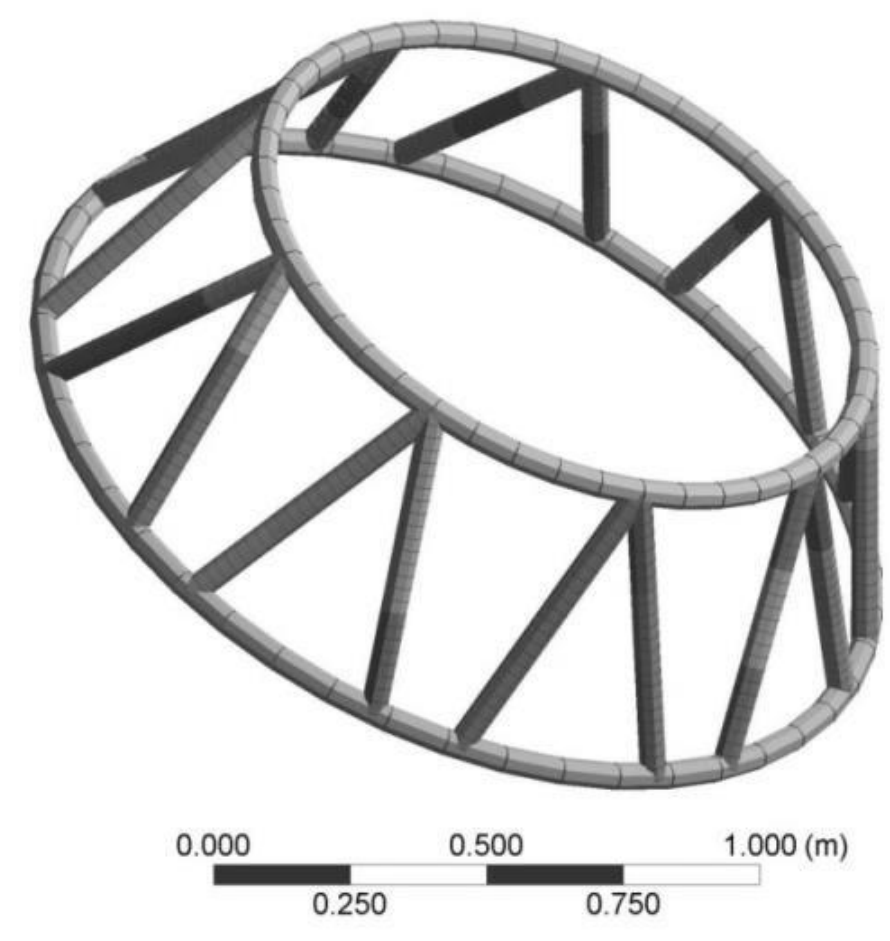

Fig. 2. Proposed satellite smart adapter with 18 active struts

Some of the critical properties of the smart adapter are listed in Table 1.

TABLE 1
SMART ADAPTER CRITICAL PROPERTIES

| Critical Properties | Value |
| --- | --- |
| Volume | 3.7881e-002 m³ |
| Satellite interface | 1650 mm |
| Centaur interface | 1200 mm |
| Height | 500 mm |
| Circular cross-section | 50 mm |

## III. FABRICATION OF THE PIEZOELECTRIC STACK ACTUATORS

Piezoelectric stack actuators are made of the piezoelectric layers which connected mechanically in series and electrically in parallel. Each piezoelectric layer is prepared from fine powders of the component metal oxides mixed in specific proportions. To form a uniform powder, the mixture is then heated and then mixed with an organic binder to form into specific circular shape. The layers will be heated for a specific time under a predetermined temperature so that the powder particles sinter and the material acquire a dense crystalline structure. The piezoelectric and the electrode materials are placed in a consolidated form and co-fired to produce a monolithic block of material with electrodes that are alternately connected. After co-firing, the stack is poled by raising the temperature closer to the Curie temperature and applying an electric field greater than the coercive field. The poling process or poling treatment is the alignment process of domains performed by exposing the element to a strong DC electric field at a temperature close to the Curie temperature.

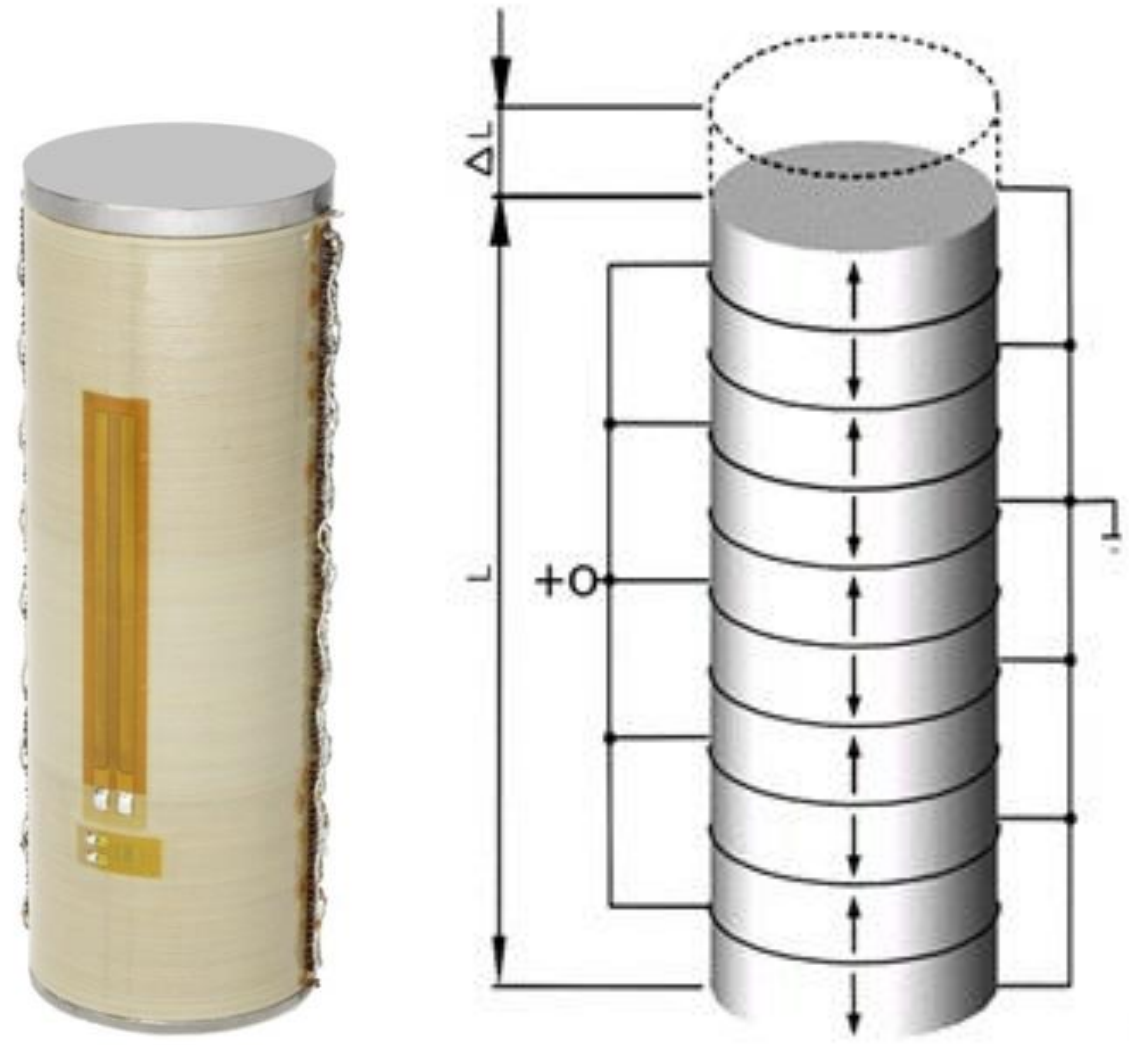

Fig. 3. Real PIC-151 piezoelectric stack actuator (left) and its schematic illustrating layers and electrodes (right)

In order to fabricate the piezoelectric stacks of PIC-151 piezo-material as shown in Fig. 3, the tapes need to be dried, cut into required dimensions, screen-printed with platinum electrode paste, and dried. The individual layers are then stacked and laminated one above the other using uniaxial stacking machine. Following the above procedure, a PIC-151 stack of height 200 mm is fabricated.

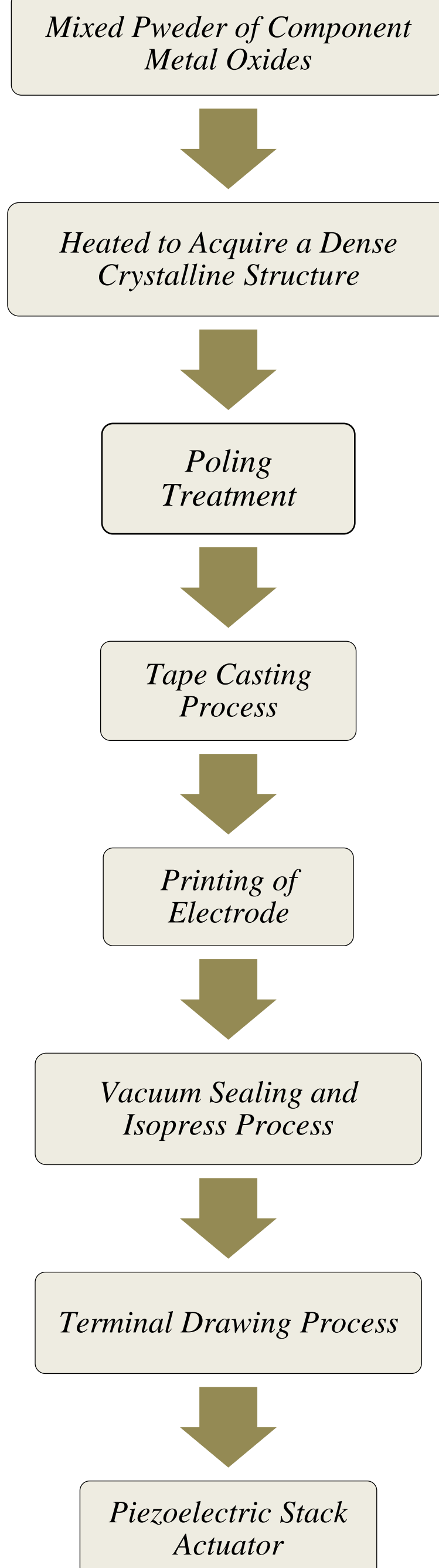

Fig. 4. Flowchart for fabrication process of piezoelectric stack actuator

## IV. DYNAMIC MODEL FORMULATION

The fundamental component of the active strut is the piezoelectric stack actuator which is made of piezoelectric material layers sandwiched between electrodes. The piezoelectric effect is sensitive to the orientation of electric field since the stacks are polarized uniaxially along their thickness to achieve maximum actuator displacement. Stack actuators are formed by assembling several wafer elements in series mechanically, connecting the electrodes result in change of the length of all layers in the same direction. In this study, conventional stacks configuration is replaced and considered as a single uniform piezo-stack actuator presented in Fig. 3 which also in turn, will enable to neglect the nonlinear material effects. The virtual term in this study is proposed to define displacement and voltage of the piezo-stack actuator (PSA) which are written in Eqs. (1) and (2);

$$\Delta l_{virtual_{PSA}} = d_{33}\, U_{virtual_{PSA}} \tag{1}$$

$$U_{virtual_{PSA}} = n\, U \tag{2}$$

where $d_{33}$ is the strain coefficient [m/V], $n$ is the number of piezoelectric layers and $U$ is the operating voltage.

The constitutive equations describing the properties of piezoelectric materials are presented in the Eqs. (3) and (4). These equations are based on assumptions that general strain in the piezo-actuator is equal to the sum of the induced mechanical strain and this mechanical strain is also due to mechanical stress and controlled strain stimulation. In the presented constitutive equations below, linear material behavior has been considered.

$$\sigma_{ij} = C_{ij}^{E} \varepsilon_i + e_{ij} E_m \tag{3}$$

$$D_m = e_{ij} \varepsilon_i + \epsilon_{ik}^{\sigma} E_k \tag{4}$$

Applying D'Alembert's principle of virtual displacements for deriving the equation of motion;

$$\int_V \left(\rho \delta u_i \ddot{u}_i + \delta \varepsilon_{ij} \sigma_{ij}\right) dV = \int_V \delta u_i b_i dV + \int_A \delta u_i t_i dA \tag{5}$$

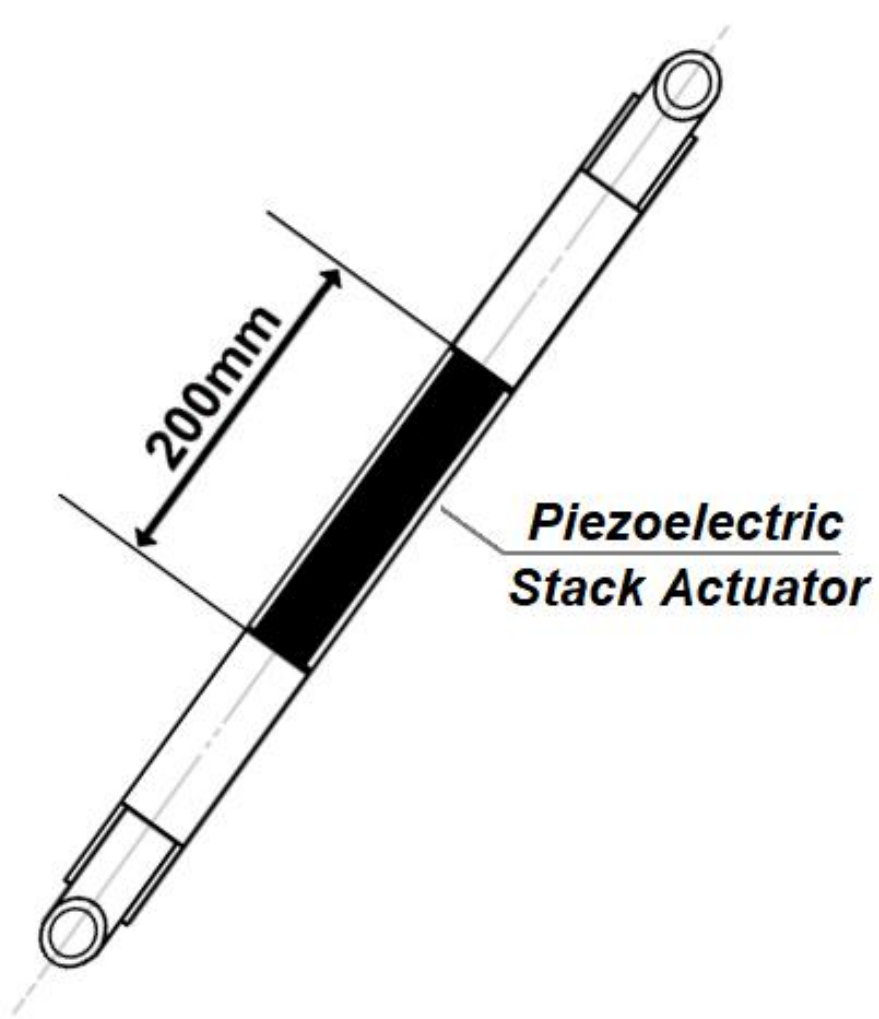

Fig. 5. Active strut configuration of satellite smart adapter

where $A$ is the cross-section of the piezo-stack actuator, $V$ is the volume, $\rho$ is the mass density, $\delta u_i$ is the virtual displacement, $\ddot{u}_i$ is the acceleration, $\delta \varepsilon_{ij}$ is the virtual strain, $\sigma_{ij}$ is the stress, $b_i$ is the volume force and $t_i$ is the traction. Also, electric flux conversion is expressed as;

$$\int_V \delta E_m D_m dV + \int_V \delta \varphi q_V dV + \int_A \delta \varphi q_A dA = 0 \tag{6}$$

where $D_m$ is the electric displacement, $\delta E_m$is the virtual electric field, $q_V$ is the charge per volume, $q_A$ is the charge per area and $\delta \varphi$ is the virtual electric potential. Using superposition for Eqs. (8) and (9) conclude the following equation,

$$\int_V \left(\rho \delta u_i \ddot{u}_i + \delta \varepsilon_{ij} \sigma_{ij}\right) dV - \int_V \delta E_m\, D_m\, dV = \int_V \delta u_i b_i dV + \int_A \delta u_i t_i dA \int_V \delta \varphi q_V dV + \int_A \delta \varphi q_A dA \tag{7}$$

where each symbol has described in nomenclature, substituting Eqs. (3) and (4) in equation (7) will result in the coupled electromechanical virtual work principle for PSA which is written as;

$$\begin{aligned} &\int_V \rho \delta u_i\, \ddot{u}_i d dV + \int_V \delta \varepsilon_{ij}\, C_{ij}^{E} \varepsilon_i dV - \int_V \delta \varepsilon_{ij} e_{ij}\, E_m dV \\ &- \int_V \delta E_m\, e_{ij} \varepsilon_i\, d - \int_V \delta E_m \epsilon_{ik}^{\sigma} E_k dV \\ &\qquad = \int_V \delta u_i b_i dV + \int_A \delta u_i t_i dA \\ &\qquad + \int_V \delta \varphi q_V dV + \int_A \delta \varphi q_A dA \end{aligned} \tag{8}$$

Using the equation (8) and finite element approximation, the consistent mass matrix, the stiffness matrix and the load vector have been calculated which are presented in Eqs. (9-11) respectively.

$$M^e = \frac{\rho A L}{30} \begin{bmatrix} 4 & 2 & -1000 \\ 2 & 16 & 2000 \\ -1 & -2 & 4000 \\ 0 & 0 & 0000 \\ 0 & 0 & 0000 \\ 0 & 0 & 0000 \end{bmatrix} \tag{9}$$

$$K^e = \frac{A}{12L} \begin{bmatrix} 7E & -8E & 1E & 7e & -8e & 1e \\ -8E & 16E & -8E & -8e & 16e & -8e \\ 1E & -8E & 7E & 1e & -8e & 7e \\ 7e & -8e & 1e & -7\xi & 8\xi & -1\xi \\ -8e & 16e & -8e & 8\xi & -16\xi & 8\xi \\ 1e & -8e & 7e & -1\xi & 8\xi & -7\xi \end{bmatrix} \tag{10}$$

$$F^e = \begin{bmatrix} N^i \\ N^j \\ N^k \\ q^i \\ q^j \\ q^k \end{bmatrix} \tag{11}$$

## V. Simulation And Control

### A. Modal Analysis

The increase of complexity in dynamic systems has made it extremely difficult and time-consuming to express the overall mathematical models with partial differential equations (PDEs), this condition is also true for the proposed smart adapter system. In this study, an efficient technique for extraction of a mathematical model of the complex smart adapter system has been utilized. Since conventional classical methods for derivation of mathematical model undergo computational error and show less accuracy, computer-aided finite element modal analysis was performed to overcome this problem. Modal analysis refers to the process of characterizing the inherent properties of smart adapter dynamics in forms of natural frequencies, damping and mode shapes. The resulting data from the modal analysis has enabled to calculate a mathematical model for the dynamic behavior of the overall system. In Table 2 the first three mode shapes of the smart adapter are presented.

### B. Control Design

Fuzzy logic controller which has found numerous application in active vibration suppression investigations [25-33] has been used in this study for active vibration control of the proposed smart adapter system. The aim is to design a closed-loop acceleration feedbacked system which utilizes triangular configured membership functions.

The fundamental of fuzzy logic control scheme consists of four main processes shown in Fig. 4;

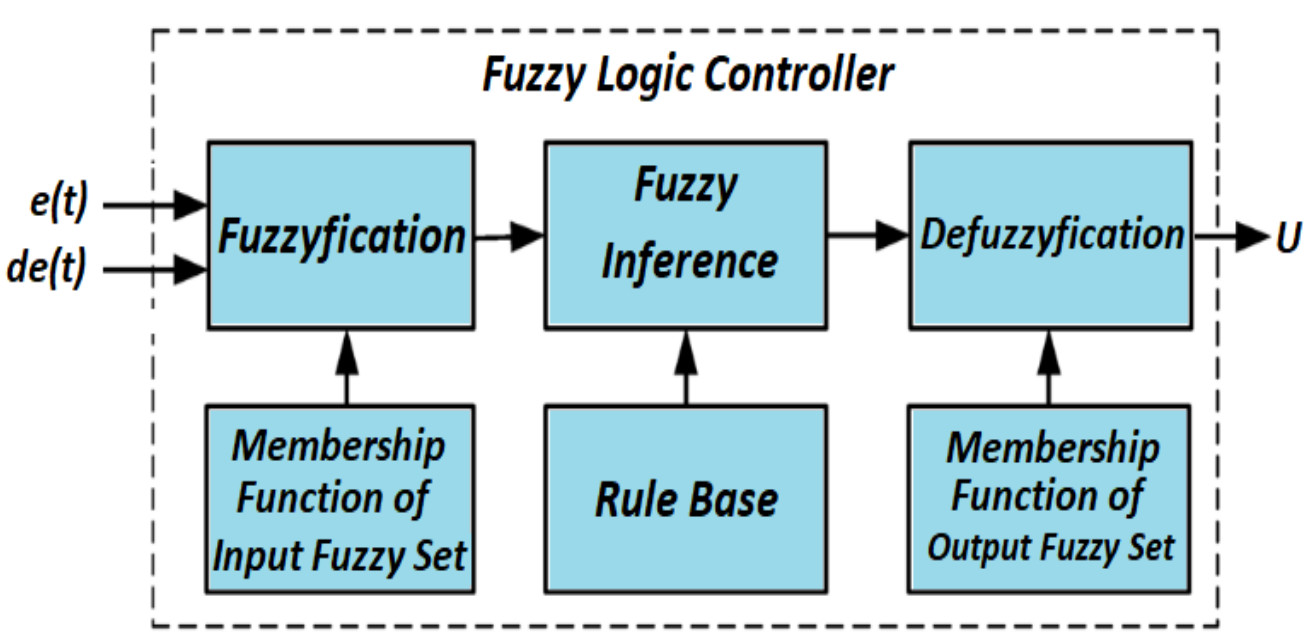

Fig. 6. Diagram of fuzzy logic controller

- **Fuzzification**: is the first step in designing a fuzzy controller which also sometimes called the parameters interpretation interface. The $e(t)$ and $de(t)$ are the most prevalent input signals which are presenting voltage signal and its derivative respectively. The output is the actuation voltage signal $u$ which is sent to piezo-stack actuators. High control efficiency is gained by applying membership functions (MFs) which require implementation of linguistic synthesis with a fuzzy logic controller. The scope of defining the MFs is to replace the control variables with control linguistic levels (CLLs).

***NOTE:*** In this study, five CLLs representing different control variable are listed as: **LP** Large Positive, **P** Positive, **Z** Zero, **N** Negative, **LN** – Large Negative.

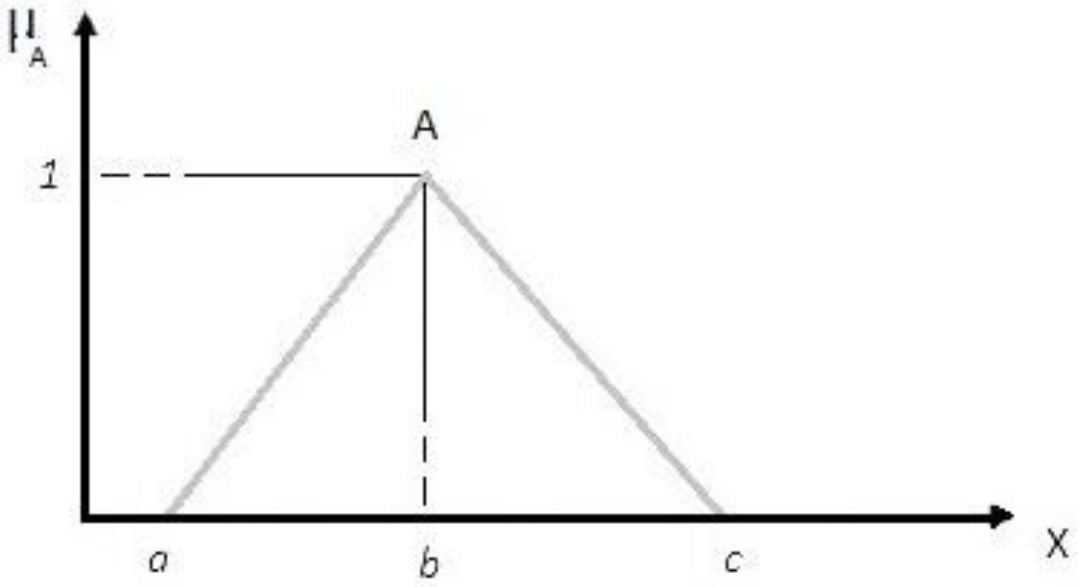

Fig. 7. Triangular MF

The schematic of the triangular configured membership function is illustrated in Fig. 5, the main aim would be to find optimum tunned values for $a$, $b$ and $c$ which are described as:

$$\mu_A(x) = \begin{cases} 0, & x \leq a \\ \dfrac{x-a}{b-a}, & a < x \leq b \\ \dfrac{c-x}{c-b}, & b < x \leq c \\ 0, & x > a \end{cases} \qquad (12)$$

The fuzzy variables e and de membership functions which are illustrated in Fig. 6 are the key for replacement process of control variables to CLLs mentioned earlier.

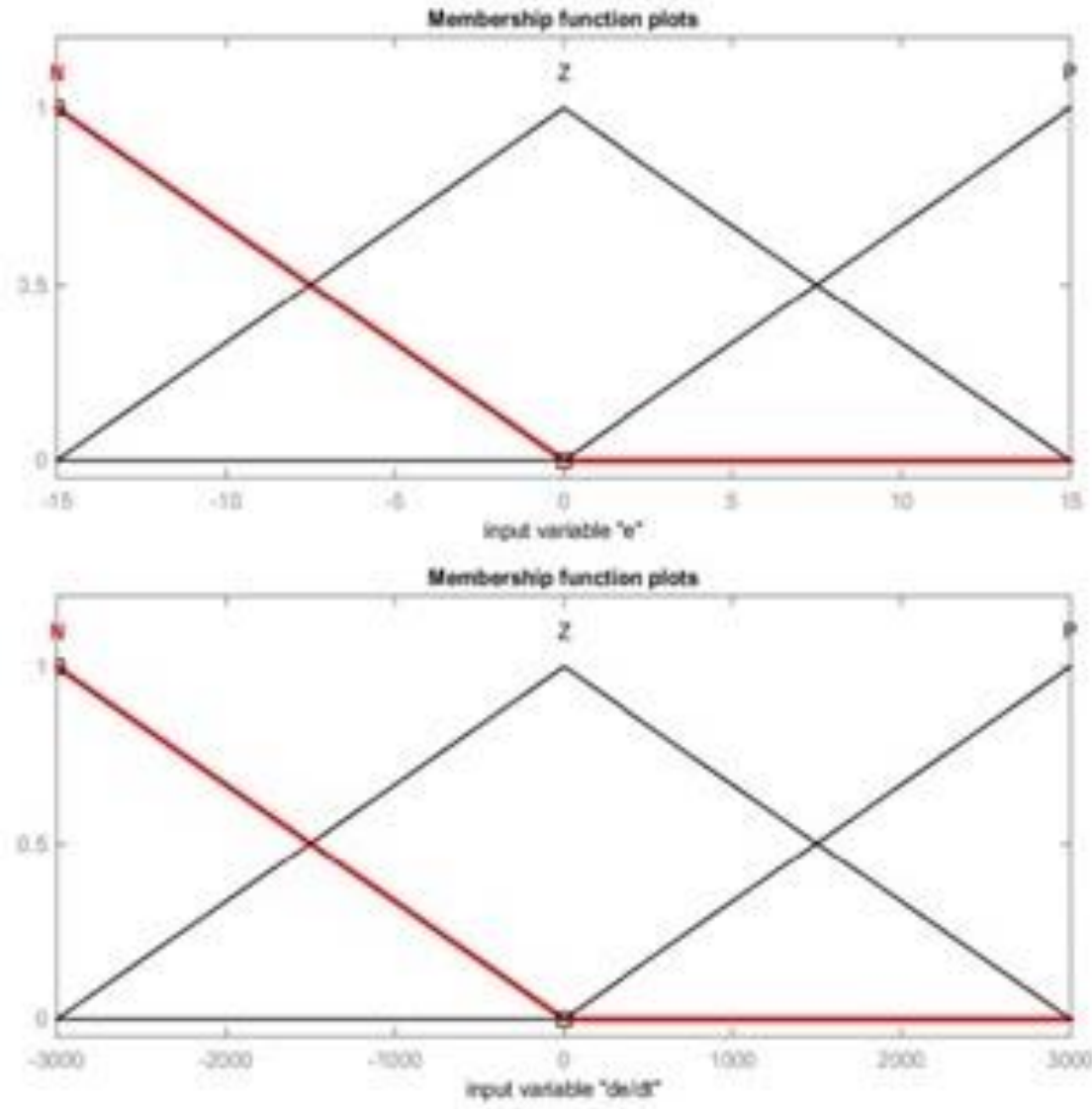

Fig. 8. Fuzzy MFs (top $e$, bottom $de$)

TABLE 2
SMART ADAPTER MODAL ANALYSIS

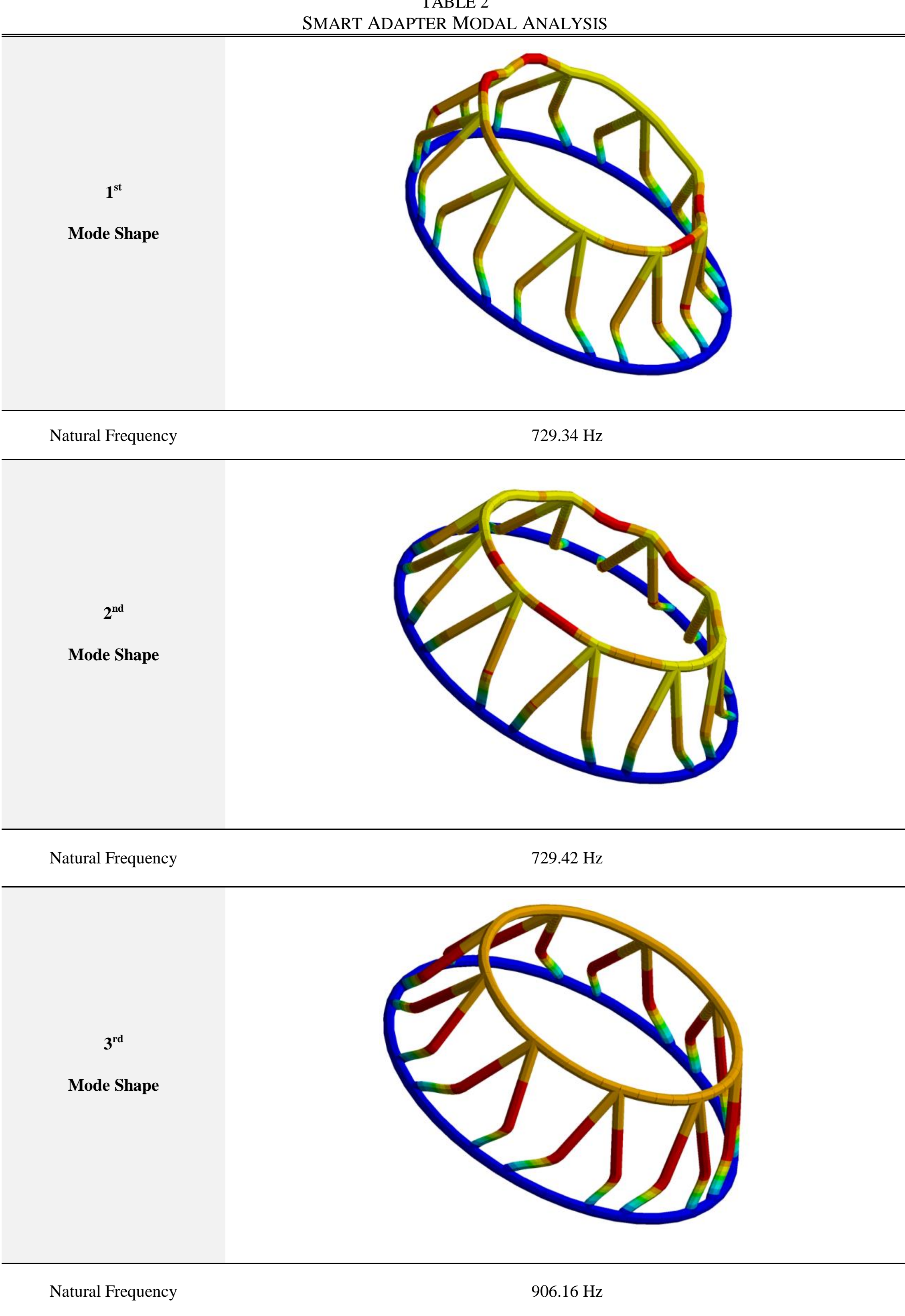

| **1st Mode Shape** | |
| --- | --- |
| Natural Frequency | 729.34 Hz |
| **2nd Mode Shape** | |
| Natural Frequency | 729.42 Hz |
| **3rd Mode Shape** | |
| Natural Frequency | 906.16 Hz |

- **Fuzzy inference and rule base:** are paying the critical role in controller efficiency. Fuzzy rules have been the main tool for expressing intended objectives in fuzzy logic systems since there is not a unique type of fuzzy rule or fuzzy logic thus it is obvious that defining control rules demands designer experience. These rules are based on IF/THEN format which offers a convenient way for expressing intended objectives. In this study, Conjunctive Antecedents Compound Rules (CACRs) format has been applied to the fuzzy control rules expression which is defined by:

$$\text{IF } e \text{ is } A_1^n \text{ AND } de \text{ is } A_2^n \text{ THEN } u \text{ is } B^n \tag{13}$$

where $A_1^n, A_2^n$ and $B^n$ are the CLLs defined by fuzzy sets on the ranges of inputs and output. The nine inferred control rules for smart adapter active vibration control are presented in Table 3.

TABLE 3
INFERENCE RULES

| de \ e | N | Z | P |
|---|---|---|---|
| **N** | *LP* | *P* | *Z* |
| **Z** | *P* | *Z* | *N* |
| **P** | *Z* | *N* | *LN* |

- **Defuzzification:** refers to the process of converting the membership degrees of CLLs output into crisp numerical values since the generated fuzzy results could not be used for further applications, this might also be expressed as rounding off process. Defuzzification procedure is performed by several mathematical methods like Center of Area (CoA), modified Center of Area (mCoA), Center of Sums (CoS), Center of Maximum (CoM) and Mean of Maximum (MoM). Selecting a defuzzification method depends on the context of the design which scoped to be calculated with the fuzzy controller. The most widely applied defuzzification technique is CoA which also referred as the Center of Gravity (CoG) has been performed in this study, the fuzzy controller evaluates the area under the scaled membership functions, then with utilizing the Eq. (11) finds its geometric center.

$$CoA = \frac{\int_{x_{min}}^{x_{max}} f(x)xdx}{\int_{x_{min}}^{x_{max}} f(x)dx} \tag{14}$$

where $x$ is the value of CLLs, $x_{max}$ and $x_{min}$ also represenrs their domains.

## VI. RESULTS AND DISCUSSION

In following figures, controlled responses of the smart adapter system are presented in the time-acceleration domains.

In the Figures 7, 8 and 9, responses of the smart adapter have been presented in the time-acceleration domains for step inputs of 15, 25 and 35 $m/_{S^2}$ respectively. The figures illustrate active and passive responses of the smart adapter system. The red colored signals are the passive responses of the system which mean that the smart adapter has surpressed vibrations when the controller was off. Also, the blue colored signals are the active responses of the smart adapter system in case when the controller was on.

In the Figures 10, 11 and 12 response signals of the smart adapter have been presented in the time-acceleration domains for sinusoidal inputs of 15, 20 and 35 $m/_{S^2}$ with 100, 125 and 150 Hz. frequencies respectively. The mentioned figures illustrate active and passive responses of the smart adapter system along with input signals colored in black. The red colored signals are the passive responses of the system which mean that the smart adapter has surpressed vibrations when the controller was off. Also, the blue colored signals are the active responses of the smart adapter system in case when the controller was on.

The results show that applying fuzzy logic control has led to achieve reliable and robust performance of the smart adapter system. Also, analysis of the results for active vibration control signals which are presented in Figures 7, 8, 9, 10, 11 and 12 using membership functions shown in Figure 6 have inferred the following outcomes:

I. Analysing results which have applied mentioned fuzzy inference rules, one knows that the designed fuzzy controller with triangular membership functions for fuzzy sets $e(t)$, $de(t)$ and $U$ can suppress the amplitude of residual vibration quite well and rapid.

II. Comparing the results shown in Figures 10, 11 and 12 which present different sinusoidal input frequencies and amplitudes, one knows that the designed fuzzy controller with triangular membership functions for fuzzy sets $e(t)$, $de(t)$ and $U$ will not change the control performance with increasing the frequency.

III. Comparing the results of system response for step and sinusoidal input, it can be seen that the control performances are not be altered using the same fuzzy controller with triangular membership functions for fuzzy sets $e(t)$, $de(t)$ and $U$.

IV. Comparing the results using active and passive system signals, it can be seen that the large amplitude launch vehicle vibrations are attenuated effectively on both active and passive modes.

The control results and their analyses demonstrate that the designed pure fuzzy logic controller can suppress the launch vehicle vibrations successfully.

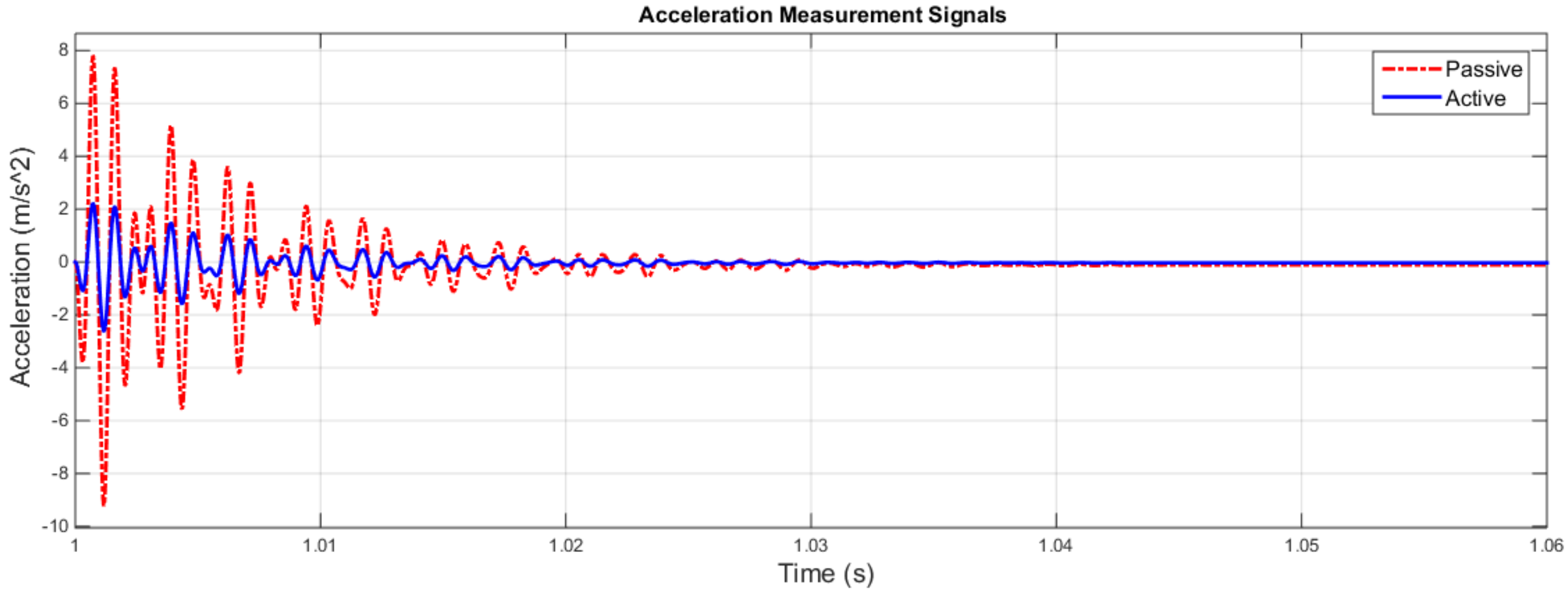

Fig. 9. Acceleration response signals for 1.5 $g$ step input

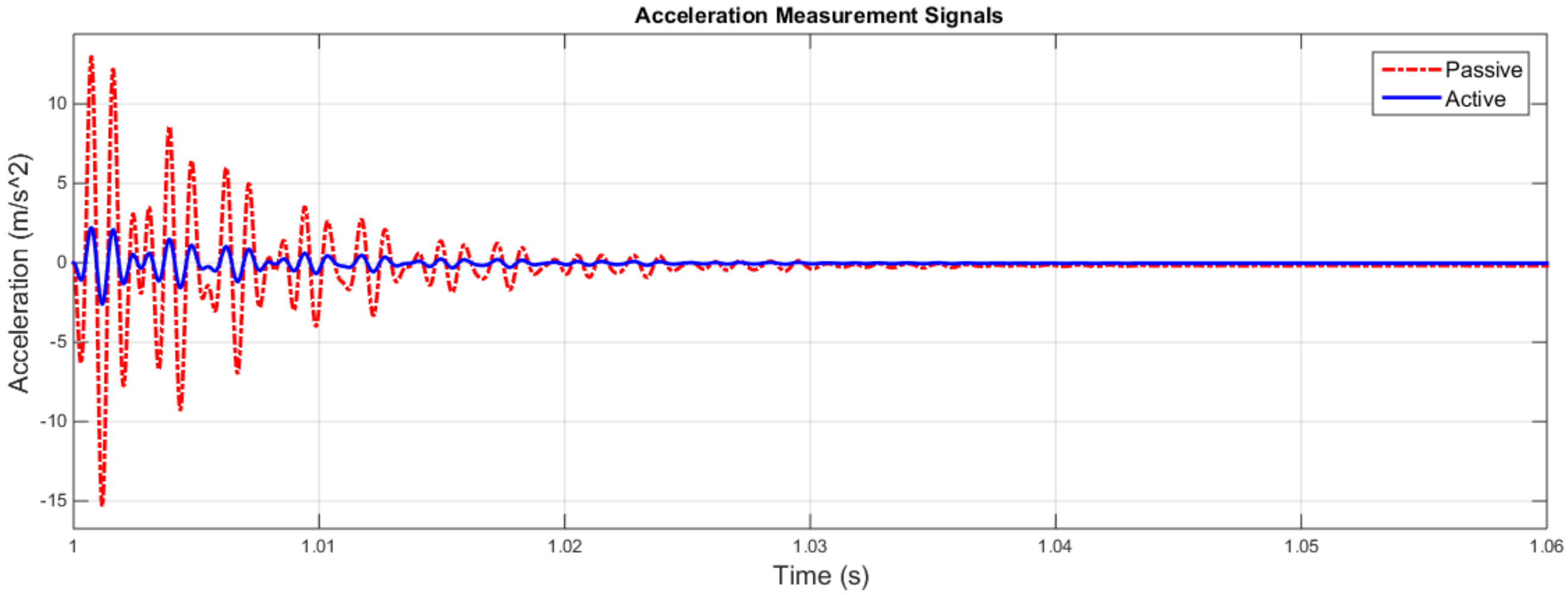

Fig. 10. Acceleration response signals for 2.5 $g$ step input

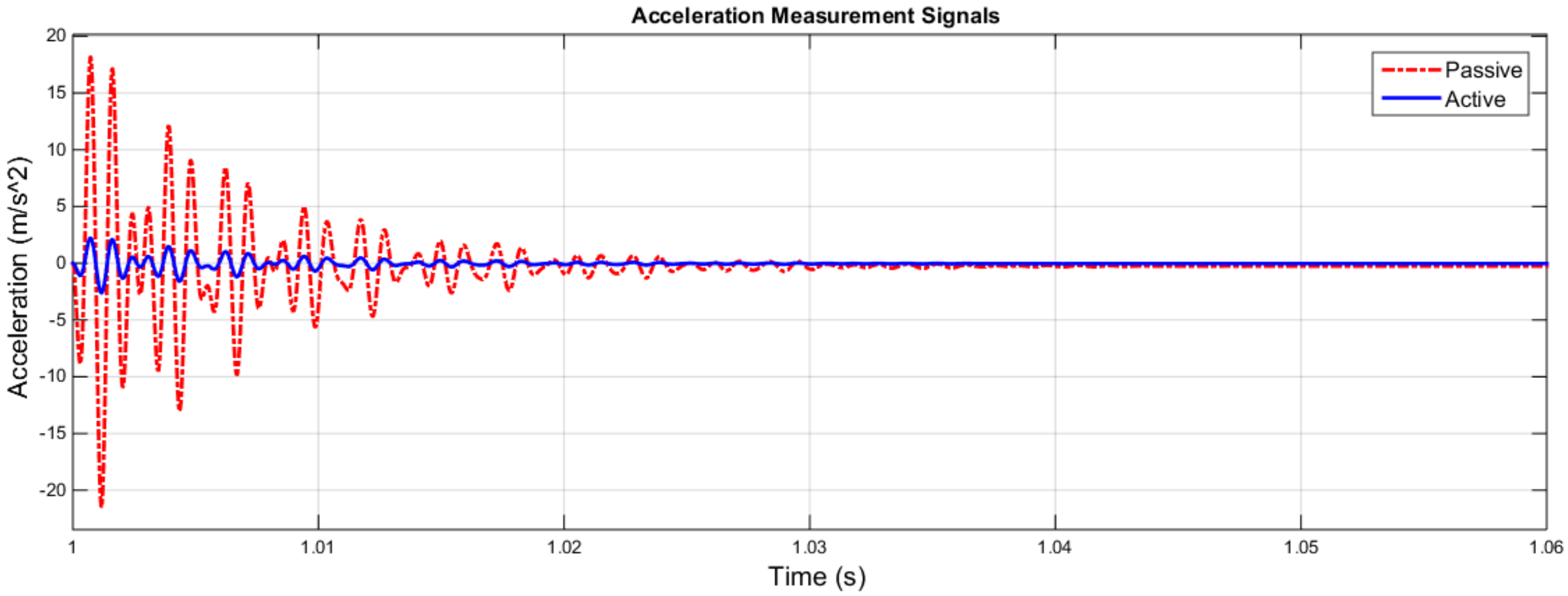

Fig. 11. Acceleration response signals for 3.5 $g$ step input

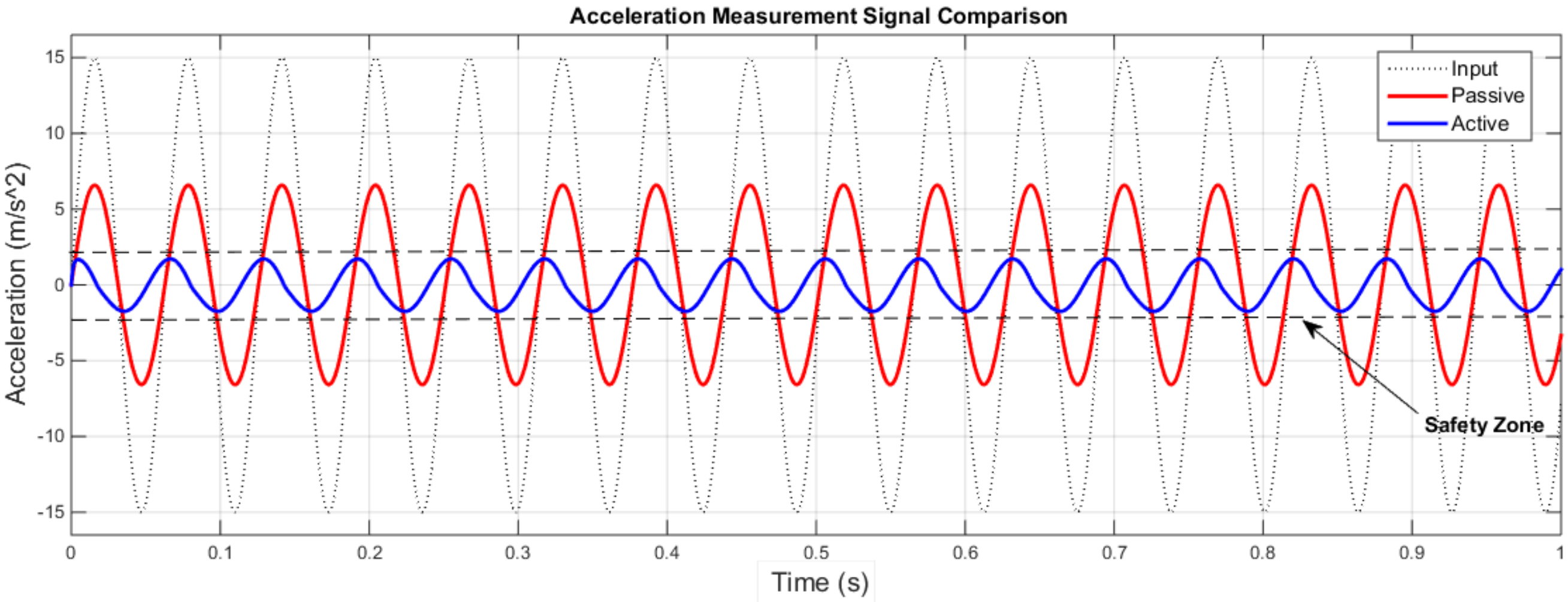

Fig. 12. Acceleration response signals comparsion for sinusoidal input (1.5$g$, 100Hz.)

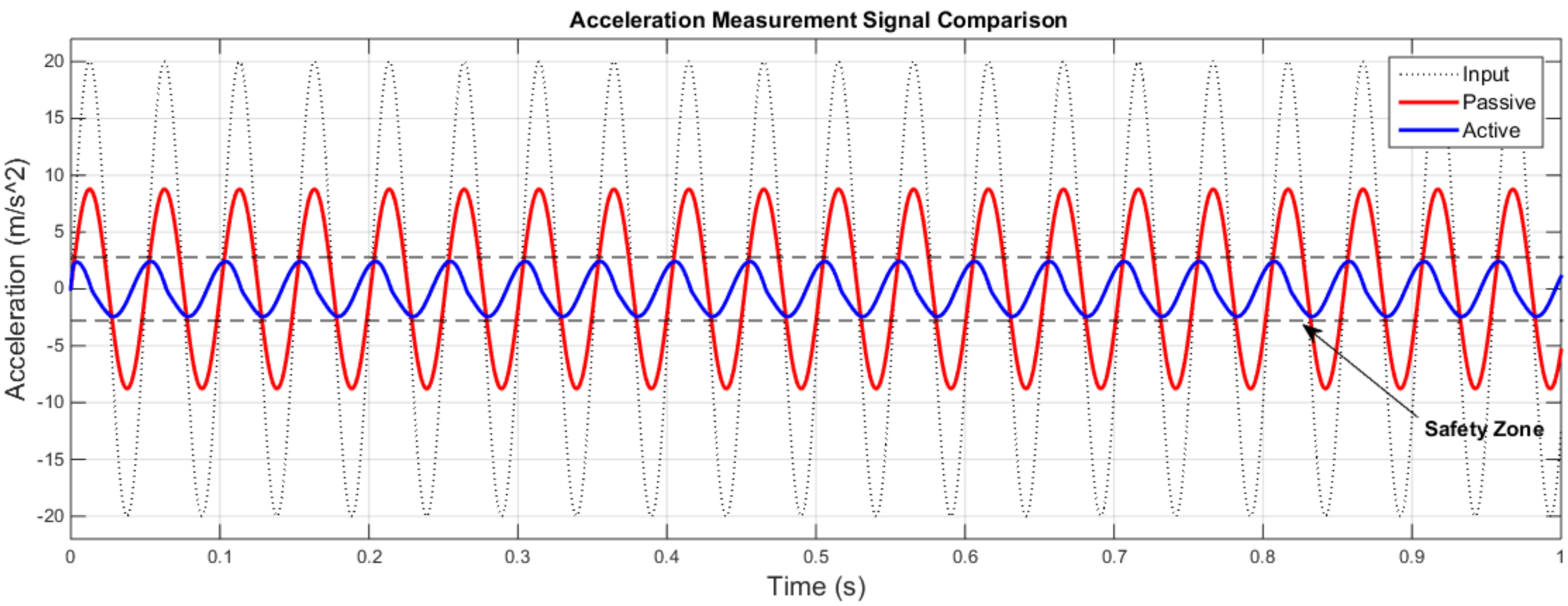

Fig. 13. Acceleration response signals comparsion for sinusoidal input (2 $g$, 125 Hz.)

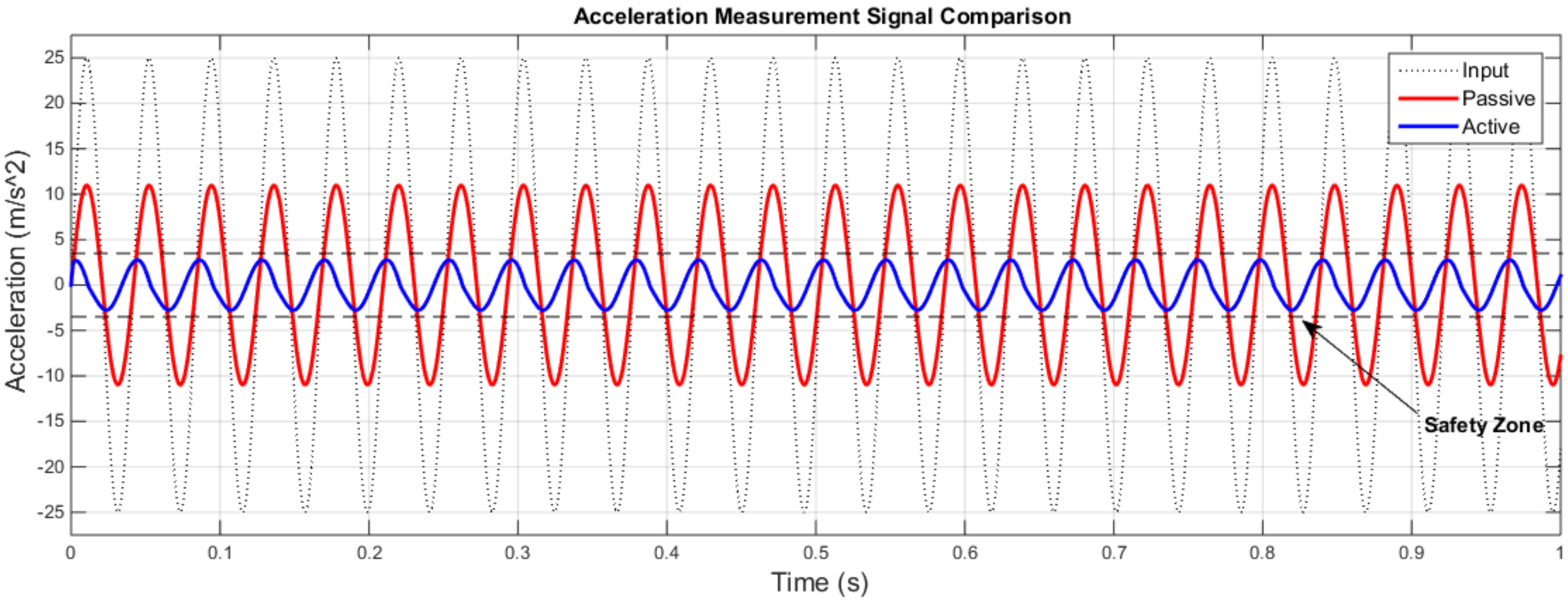

Fig. 14. Acceleration response signals comparsion for sinusoidal input (2.5$g$, 150Hz.)

## VII. Conclusions

The main aim of this study is to propose a satellite smart adapter for active vibration control of launch vehicle on satellite, the proposed smart adapter benefits from the most effective and efficient space actuator as elements of its intelligent structure. Comprehensive conceptual design of satellite smart adapter is presented to indicate and illustrate the design parameters and requirements along with design philosophy applied in this study. It is worth to note that the selection of smart material has been based on reliability and durability criterions to ensure successful functionality of the proposed system, also fabrication process of the proposed piezoelectric stack actuator is discussed in detail. The coupled electromechanical virtual work equation for a piezoelectric stack actuator in each active strut is drived by applying D'Alembert's principle to calculate the consistent mass matrix, the stiffness matrix and the load vector using finite element approximation. Modal analysis is performed to characterize the inherent properties of the smart adapter and extraction of a mathematical model of the system. Active vibration control analysis was conducted using fuzzy control with triangular membership functions. The control results demonstrate that the designed pure fuzzy logic controller can suppress the launch vehicle large amplitude vibrations successfully. Comparing the results of the system response for the step and sinusoidal inputs, it can be proved that the control performances are not be altered using the same fuzzy controller with triangular membership functions and increasing the input frequency will not change the control performance. Finally, it is concluded that the proposed satellite smart adapter configuration which benefits the piezoelectric stack actuator as elements of its 18 active struts has high strength and shows excellent robustness and effectiveness in vibration suppression of launch vehicle vibrations on satellite.